\documentclass[reprint, amsmath, amssymb, aps, prapplied, superscriptaddress]{revtex4-2}
\usepackage{appendix}
\usepackage{graphicx}
\usepackage[utf8]{inputenc}

\begin{document}

\author{M.\ Bia\l{}ek\normalfont\textsuperscript{$\dagger$}}
\email{marcin.bialek@epfl.ch}
\address{Institute of Physics, \'Ecole Polytechnique F\'ed\'erale de Lausanne (EPFL), 1015 Lausanne, Switzerland}
\author{J.\ Zhang}
\thanks{These two authors contributed equally to this work}
\address{Fert Beijing Institute, School of Microelectronics, Beijing Advanced Innovation Center for Big Data and Brain Computing, Beihang University, Beijing 100191, China}
\author{H.\ Yu}
\email{haiming.yu@buaa.edu.ch}
\address{Fert Beijing Institute, School of Microelectronics, Beijing Advanced Innovation Center for Big Data and Brain Computing, Beihang University, Beijing 100191, China}
\author{J.-Ph.\ Ansermet}
\address{Institute of Physics, \'Ecole Polytechnique F\'ed\'erale de Lausanne (EPFL), 1015 Lausanne, Switzerland}

\title{Strong coupling of antiferromagnetic resonance with sub-THz cavity fields}
\date{\today}

\begin{abstract} 
Strong coupling of electromagnetic cavity fields with antiferromagnetic spin waves in hematite ($\alpha$-Fe$_2$O$_3$) was achieved above room temperature.
A cube of hematite was placed in a metallic tube and transmission was measured, using a continuous-wave THz spectrometer. Spectra, collected as a function of temperature, reveal the formation of magnetic polaritons.
\end{abstract}

\maketitle

\section{Introduction}
The hybrid nature of strongly coupled light-matter states, were the dissipation rate is lower than the exchange rate (Rabi frequency) \cite{Torma14}, has attracted a lot of attention since the late 1980s \cite{Rempe87}. In the 2000s decade, strong light-matter coupling was investigated in the solid state \cite{Khitrova06}, where the coupling of light interacting with $N$ resonators increases by a factor of $\sqrt{N}$ \cite{Raizen89}. This regime might lead to device elements that may play a role in quantum devices \cite{Kasprzak06, Awschalom07, Torma14, Dovzhenko18, Kockum19, Roux20}. In the 2010s, interest turned toward magnon-photon coupling in ferromagnetic materials \cite{Schuster10, Abe11, Huebl13, Zhang14, Tabuchi14, Tabuchi15, Zhang15, Zhang16, Li19, Potts20, Lachance-Quirion20, Li20JAP, Bhoi21}, as spins benefit from a relatively low coupling to their environment \cite{Awschalom07, Niemczyk10}. Achieving strong coupling with antiferromagnets has the advantage of operating at frequencies in the terahertz (THz) range \cite{Jungwirth18, Li20Nature} that brings a lot of interesting physics unreachable in ferromagnets \cite{Li20PRL, Reitz20, Ghosh21}. There is a large variety of antiferromagnets available, and many of them show magnetic ordering above room temperature. The absence of stray fields could in principle allow for extremely dense packing of elements \cite{Hoffman15, Jungwirth18}. However, there are reports only on a few examples of weak magnon-photon coupling in antiferromagnets \cite{Li11, Bialek20}. Strong coupling was only achieved at low frequencies \cite{Everts20}, in bulk samples \cite{Grishunin18, Shi20} or via an indirect coupling \cite{Li18, Sivarajah19}, while direct strong cavity-magnon coupling had not been shown so far \cite{Hu20}.
One of the technical difficulties is constructing high-frequency cavities that are well coupled to wave guides. Here, we present a relatively simple implementation that allows strong coupling of the antiferromagnetic resonance and cavity electromagnetic modes.

In this letter, we show strong coupling between electromagnetic fields of a 3-dimensional cavity and spin waves in hematite ($\alpha$-Fe$_2$O$_{3}$), a very common room-temperature antiferromagnet. 
Hematite crystallizes in approximately hexagonal structure with space group R$\overline{3}$c. Precise measurements show that the actual symmetry is monoclinic C2/c or C2'/c' \cite{Przenioslo14}. Below the N\'eel temperature $T_N\approx955$~K \cite{Morin50}, the Fe$^{3+}$ magnetic moments orient in an antiferromagnetic order. Above the spin-reorientation transition (Morin phase transition) at about $T_M\approx260$~K \cite{Aleksandrov85}, the superexchange Dzialoshinskii-Moriya interaction leads to a canting of the two sublattices that gives rise to a net magnetisation $\mathbf{m}$ \cite{DZYALOSHINSKY58}, i.e. making this material a weak ferromagnet. 
Owing to the spin canting, the antiferromagnetic resonance has two modes, one of which is at a high frequency. This quasi-antiferromagnetic resonance (qAFMR) mode is excited by dynamical magnetic field $\mathbf{h}$ parallel to the magnetization ($\mathbf{h}\parallel \mathbf{m}$). At temperatures above room temperature the width of qAFMR is about 1~GHz only, and its frequency rises sharply with temperature \cite{Aleksandrov85}. 

\section{Experimental}
\begin{figure}
\begin{center}
\includegraphics[width=0.95\linewidth]{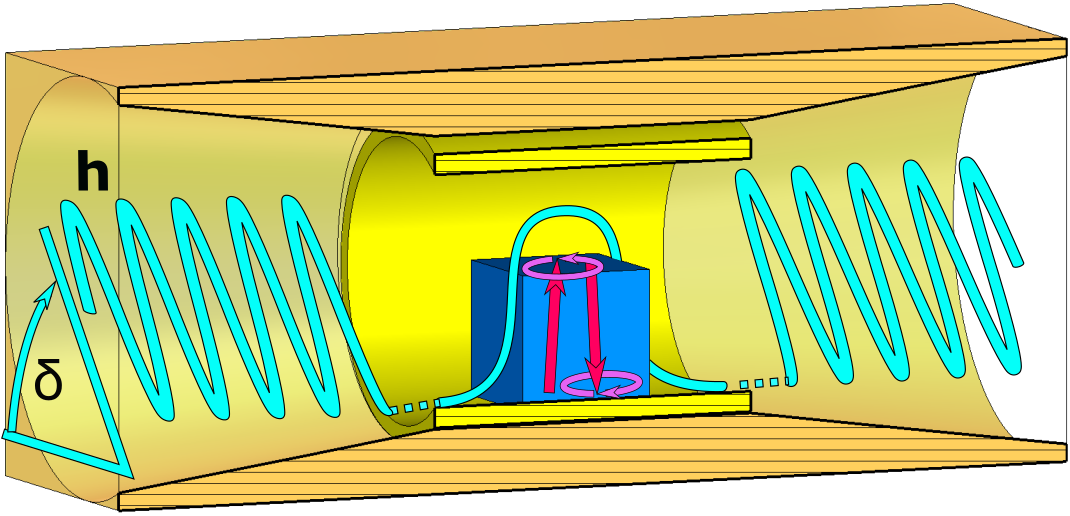}
\caption{{\label{setup}
Plane electromagnetic wave (blue wiggles) is focused into a cylinder cavity containing a hematite cube. The cavity magnetic field (blue curve) interacts with the resonance (violet circles) of two antiferromagnetic sublattices (red arrows). Measured: transmitted linear polarization parallel to that of the incident beam.}}
\end{center}
\end{figure}
Thanks to the development of frequency extenders for vector network analyzers (VNA), continuous-wave spectroscopic measurements up to 1.5~THz can be rapidly conducted with a high frequency resolution and with a very high dynamic range \cite{Caspers16, Bialek18, Bialek19, Bialek20, Zhang20}. 

Out of a bulk natural single crystal, we cut a cube of $l_s=0.2$~mm edge length, with faces in the $a$, $b$ and $c$ crystalline directions. The cube was placed inside a tube of $l_c=2$~mm length and $2r_c=0.58$~mm internal diameter, cut out of a gold-plated stainless steel needle. The needle was wrapped with Teflon tape to prevent the cube from dropping out of the tube. 
The tube with the cube in it was inserted in a copper holder composed of two cones (cone angle, about 60$^{\circ}$) which focused the THz beam into the cavity (Fig.~\ref{setup}). The THz beam propagated between the source, sample holder and detector in oversized metallic waveguides, 8~mm in diameter, matching the larger diameter of the cones. The holder was placed between Peltier elements that controlled its temperature, which was monitored with a K-type thermocouple inserted in a hole on its side. 
The detector measured the transmitted power and phase of the THz electric field. Temperature scans started from the highest temperature with a step of $\Delta T=-0.25$~K. This temperature step was so chosen that the frequency change was smaller than the line width of the resonance. After stabilizing the sample temperature $T$, we measured transmission as a function of radiation frequency $f$.

The source emits linearly polarized beam of THz radiation and the detector detects only radiation of linear polarization that matches its waveguide, i.e. it acts like a polarization filter. 
Rotation of the polarization plane is possible by rotating source and detector about the optical axis. We define the polarization angle $\delta$ as the angle of the radiation $h$ field with respect to the plane of the optical table (Fig.~\ref{setup}). We measured transmission in a polarization angle range of $\delta=-90^{\circ}$ to $+90^{\circ}$.

\section{Results and analysis}
\begin{figure}
\includegraphics[width=\linewidth]{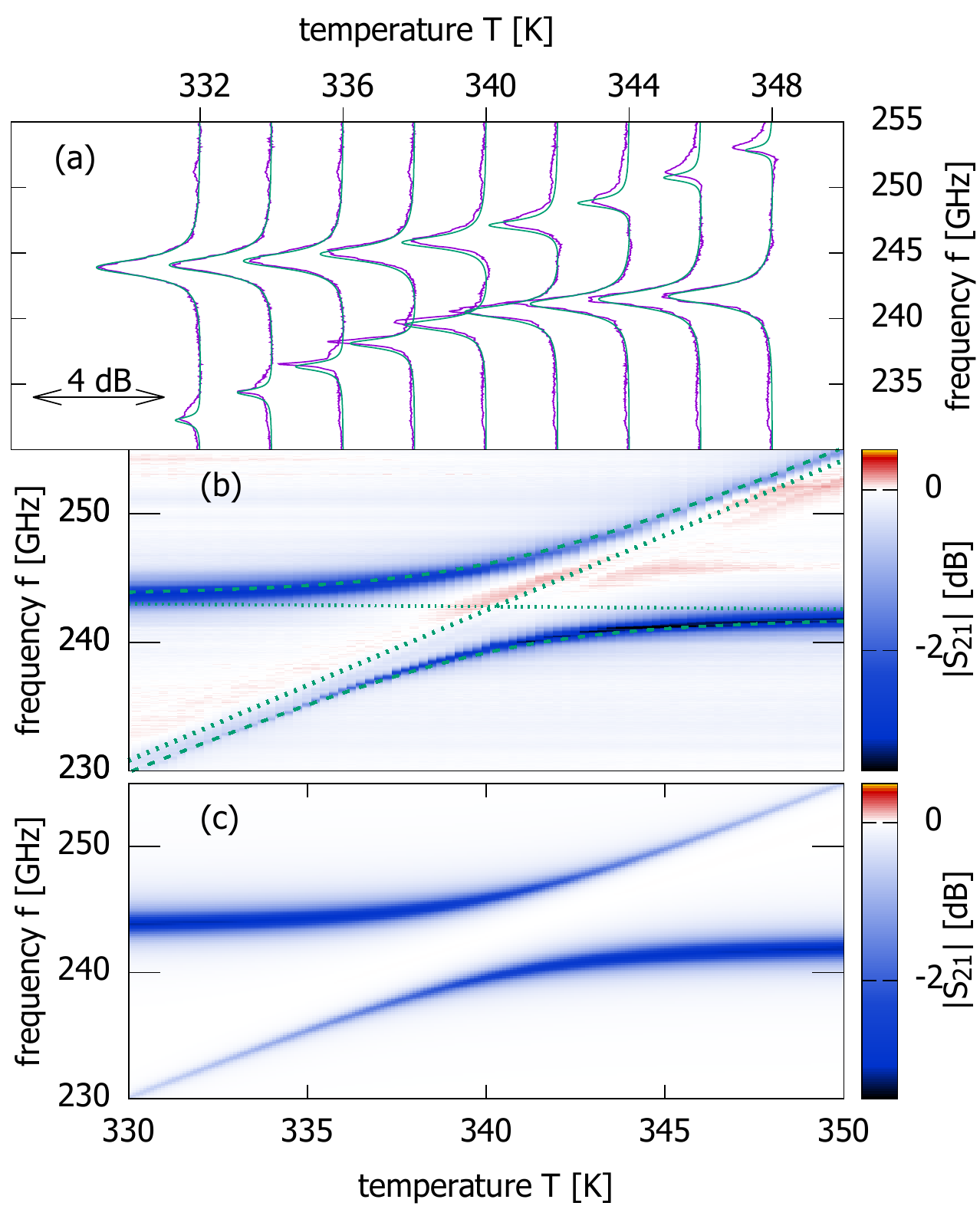}
\caption{\label{cav-res}
 Normalized transmission magnitude at polarization angle $\delta=30^{\circ}$.
 (a) Measured spectra (violet lines) and fit using Eq.\ \ref{S21-cav} (green lines). (b) Map of the spectra, green dashed lines show interacting and non-interacting modes fitted with Eq.\ \ref{anticross}. (c) Map of spectra fitted with Eq.\ \ref{S21-cav}. 
}
\end{figure}
In Fig.\ \ref{cav-res}a we can see avoided crossing of the cavity mode at $f_1\approx242.5$~GHz with the qAFMR mode, the frequency of which $f_r(T)$ rises approximately linearly with temperature (see appendix). This result was obtained at polarization angle $\delta=30^{\circ}$.
We fitted the observed spectra using the equation developed in the framework of input-output theory \cite{Schuster10, Harder16}:
\begin{equation}
    S_{21} = 1 +\frac{a_1}{i(f-f_1(T)) - \frac{\kappa_1}{2}+\frac{G_1^2}{i(f-f_r(T))-g/2}},
    \label{S21-cav}
\end{equation}
where, $a_1=(0.32-i0.02)$~GHz is a complex parameter describing the coupling of the cavity with the source and the detector (see Appendix). The parameter $f_1(T)\approx242.5$~GHz is the observed frequency of the 1st cavity mode and $\kappa_1=2.1$~GHz describes its width, $2G_1=6.2$~GHz is the minimum splitting between upper and lower polariton branches and $g=0.5$~GHz describes qAFMR width. 
With these parameters, we can determine the cooperativity factor
\begin{equation}\label{coop_factor}
C=\frac{4G_1^2}{\kappa_1 g}\approx40,
\end{equation}
that is, the square of a ratio of the Rabi splitting ($2G_1$) and the losses of the polariton states. This means that the splitting was about 6 times larger than the linewidth. According to the harmonic coupling model, the upper and lower polariton frequencies are given by \cite{Mills74, Huebl13},
\begin{equation}
f_{\pm} = \frac{1}{2}\left(f_1 + f_r \pm\sqrt{(f_1-f_r)^2+4g_{i}^2f_1}\right).
\label{anticross}
\end{equation}
The coupled mode frequencies $f_{pm}$ are drawn on Fig.\ \ref{cav-res}b as green dotted lines, with $2g_i\sqrt{f_1}=2G_1=6.2$~GHz, as well as the non-interacting mode frequencies $f_r(T)$ and $f_1(T)$.
Clearly, this prediction matches that of Eq.\ \ref{S21-cav} (Fig.\ \ref{cav-res}b). The discrepancies between fits using Eq.\ \ref{S21-cav} and data (Fig.\ \ref{cav-res}a) may arise from interactions with other cavity modes, as we discuss in further sections of this communication

\subsection{Microscopic model}
The coupling strength can be calculated using a microscopic model \cite{Huebl13}
\begin{equation}
g_i = \frac{g_s\mu_B}{2h}\sqrt{\frac{\mu_0h}{2}\rho\frac{V_s}{V_c}},
\label{coupling}
\end{equation}
where, $g_s=2$, $\rho$ is density of resonators in the hematite cube of a volume $V_s=l_s^3$ and $V_c$, the volume of the metal tube, $V_c=\pi r_c^2l_c\approx0.528$~mm$^3$.
In Eq. \ \ref{coupling}, each magnon is coupled to the electromagnetic cavity mode with a coupling strength $g_s\mu_BB_0/2h$ \cite{Soykal10}, where $B_0 = \sqrt{\mu_0h f_1/2V_c}\approx 4.4\cdot10^{-10}$~T is the magnetic component of vacuum fluctuations \cite{Niemczyk09} (here $f_1=242.5$~GHz). This is a few orders of magnitude smaller than the amplitude of the THz field in our experiment. The collective coupling strength of $N=\rho V_s$ oscillators is increased by a factor $\sqrt{N}$ \cite{Raizen89, Soykal10, Huebl13, Tabuchi14, Li18}, thus $g_i$ depends only on the oscillators density $\rho$ and the ratio of the crystal volume to the cavity volume \cite{Huebl13, Li18}, that is $V_s/V_c\approx1.51\times10^{-2}$. 

We assumed the density of oscillators in hematite is  $\rho=5\rho_{Fe}$ \cite{Flower19, Bourhill20}, where $\rho_{Fe}=3.987\times10^{28}$~m$^{-3}$ \cite{Pailhe08} and factor $5$ comes from the magnetic moment of Fe$^{3+}$ ions \cite{Shull51}. Density of iron atoms in hematite is quite high compared to that of many common antiferromagnets. Hence, it is a good material to achieve strong coupling. Taking this into account gives $g_i=1.57\times10^4$~Hz$^{1/2}$, i.e.\ the splitting is predicted to be $2G_p=2g_i \sqrt{f_1}\approx 15.4$~GHz for the cavity mode at $f_1=242.5$~GHz. The observed splitting $2G_1=6.2$ GHz is about 40\% of that value. This discrepancy results from an imperfect matching of the electromagnetic mode with the antiferromagnetic resonance in the hematite cube which is excited by fields parallel to the weak ferromagnetic moment. As confirmed in the simulation discussed below (Fig.\ \ref{mode-sim}), the cavity magnetic field in the volume of the cube does not excite all the spins with full amplitude. We calculated that in the 1st mode, the magnetic field component $h_z^2$ makes about 78\% of all the integrated magnetic field energy in the cube volume (Tab.\ \ref{modes}). The remaining difference between the predicted and the observed splitting must result from the magnetic moment of the cube being titled away from the $z$ axis. 

\subsection{Angular dependence}
\begin{figure}
\includegraphics[width=\linewidth]{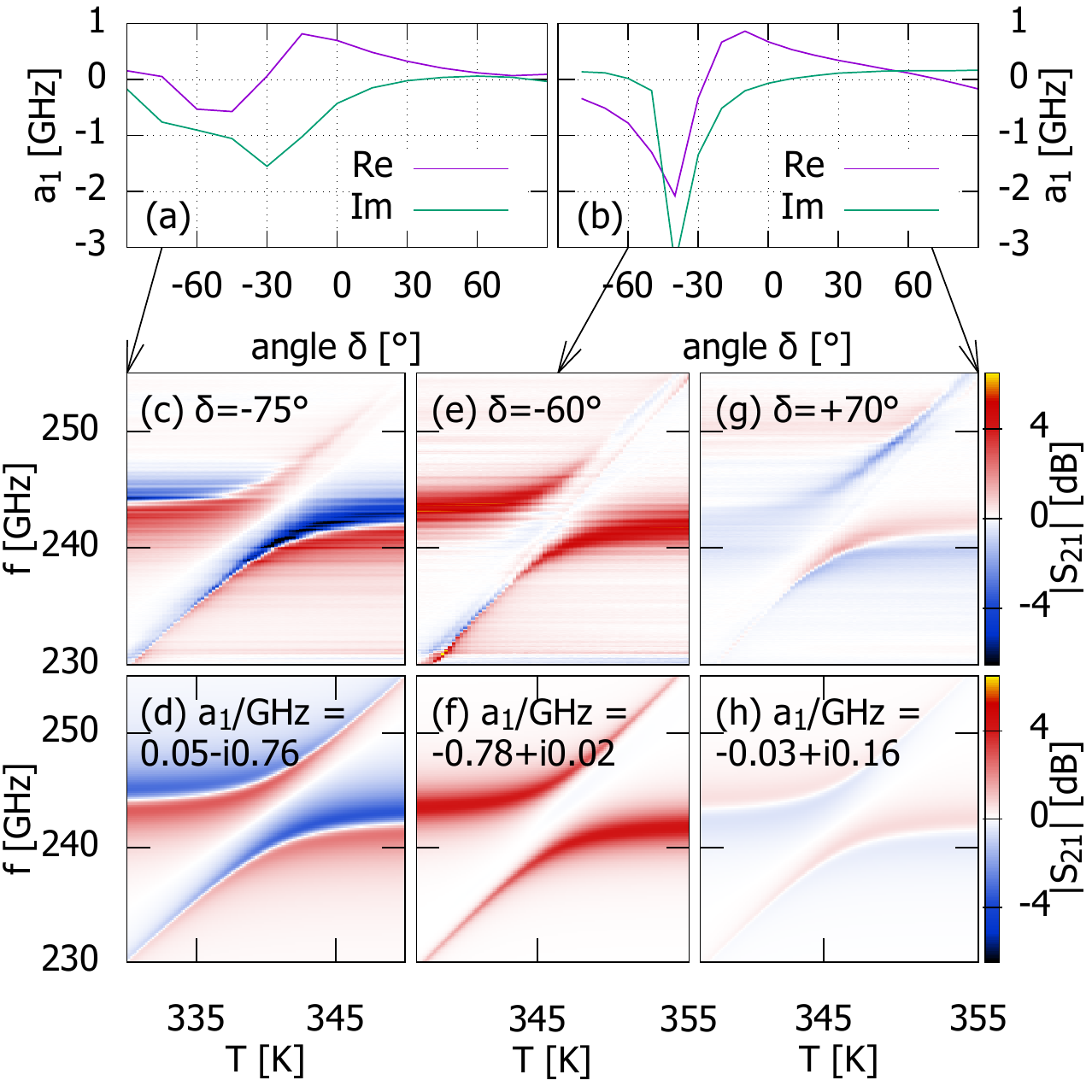}
\caption{\label{cav-delta} 
(a, b) Angular dependence of $a_1$ (Eq. \ref{S21-cav}) in two configurations of the experiment. (c) Normalized $|S_{21}|$ at $\delta=-75^{\circ}$ in (a) configuration. (e) Normalized $|S_{21}|$ at $\delta=-60^{\circ}$ and (g) $\delta=70^{\circ}$, both in configuration (b). (e, f, g) Fits using Eq.\ \ref{S21-cav}. 
}
\end{figure}
We found that the parameter $a_1$ strongly depends on the polarization angle $\delta$ (Fig.\ \ref{cav-delta}a). This parameter describes the coupling between the  cavity and the rest of the spectrometer, i.e.\ constructive or destructive interferences between the cavity and the experimental setup. The result presented in Fig.\ \ref{S21-cav}a, obtained at $\delta=30^{\circ}$, shows a minimum in transmission that is accounted for with $a_1$ having a small imaginary part and a positive real part. The dependence of $a_1$ on polarization angle (Fig.\ \ref{cav-delta}a) is not an intrinsic property of the cavity, but depends on the coupling of the cavity with the experimental setup. As a consequence, after it was rearranged, we obtained a slightly different value for $a_1$ (Fig.\ \ref{S21-cav}b).
Thus, the result obtained at $\delta=-75^{\circ}$ (Fig.\ \ref{cav-delta}c) under (a) conditions shows a dispersive lineshape that is accounted for by a negative imaginary $a_1$ (Fig.\ \ref{cav-delta}e).
In contrast, the result obtained at $\delta=-60^{\circ}$ (Fig.\ \ref{cav-delta}d) under (b) conditions shows constructive interference that is explained by an almost entirely real, negative $a_1$ (Fig.\ \ref{cav-delta}f).
Under the same conditions, at $\delta=70^{\circ}$ a weak inversed dispersive lineshape was observed (Fig.\ \ref{cav-delta}f) that is accounted for with a small positive imaginary $a_1$.

\subsection{Higher cavity modes}
Raw data for the spectra are dominated by multiple interferences that are not strongly interacting with the cavity and are weakly temperature-dependent.
We normalized both the measured spectra and fitted functions to a base frequency that is temperature-dependent and passes though the middle of the interaction. This reduced artifacts at the cavity mode frequency that would show if instead, we normalized for example to the first recorded spectrum. We could eliminate this interference background even better by calculating temperature-derivative spectra, i.e. by subtracting from one another successive spectra, thus having amplitude derivatives:
\begin{equation}
  \frac{d |S_{21}|}{d T} [dB] = \frac{20}{\Delta T}\log_{10}\frac{|S_{21}(f,T+\Delta T)|}{|S_{21}(f,T)|},
  \label{dS21dTmag}
\end{equation}
and phase derivatives:
\begin{equation}
  \frac{d(\arg{S_{21}})}{d T} = \frac{\arg S_{21}(f,T+\Delta T)-\arg S_{21}(f,T)}{\Delta T}.
  \label{dS21dTpha}
\end{equation}
\begin{figure}
\begin{center}
\includegraphics[width=\linewidth]{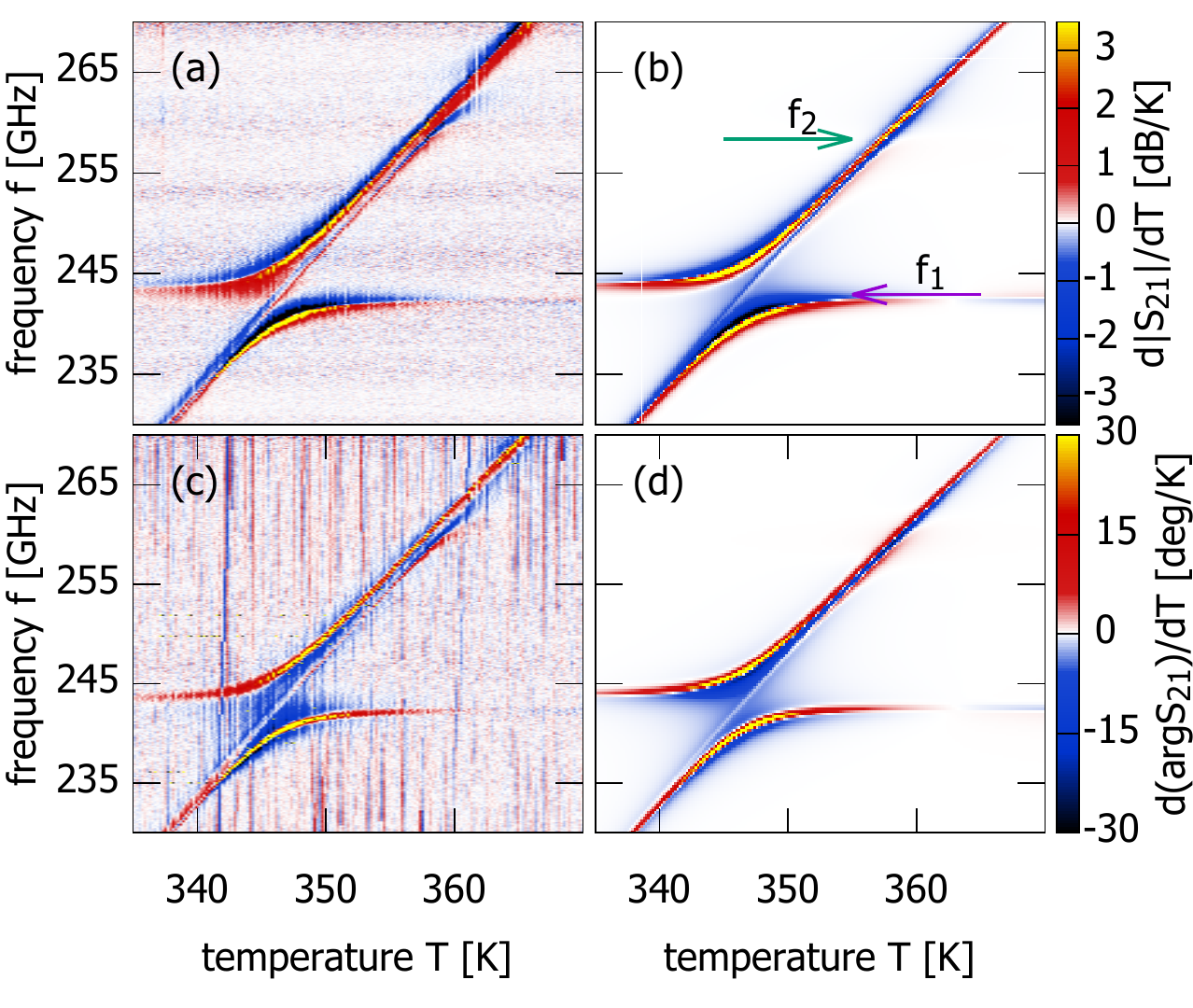}
\caption{{\label{zoom}
Temperature-derivative spectra: (a) magnitude and (c) phase. (b and d) Corresponding fits using Eq.\ \ref{S21-cav2}. }}
\end{center}
\end{figure}
The advantage of looking at temperature-differential spectra is that it is possible to see small features, like a weak middle line passing through the interaction region (Fig.\ \ref{zoom}a and c) that is almost invisible in normalized spectra. This middle line can be qualitatively explained by taking into account the 2nd cavity mode at $f_2=258.4$~GHz. We can take it into account by assuming that both modes interact independently with the magnetic resonance mode:
\begin{equation}
    S_{21} = 1+\sum_{j=1}^2\frac{a_j}{i(f-f_j(T)) - \frac{\kappa_j}{2}+\frac{G_j^2}{i(f-f_r(T))-g/2}},
    \label{S21-cav2}
\end{equation}
In Fig. \ref{zoom}b we present temperature-differential of Eq.\ \ref{S21-cav2}, where we put $a_2=(4.5-i1.1)$~GHz, $G_2=0.56$~GHz and $\kappa_2=14.7$~GHz. 
These values are consistent with our idea of a weak interaction with the 2nd cavity mode that has a lower quality factor that the 1st mode. This simple model is not fully sufficient above 250~GHz. We think that a more precise estimate might require taking into account other weakly interacting cavity modes or interaction between cavity modes that are mediated by the magnetic resonance. We found that the observed phase temperature-differential spectra (Fig. \ref{zoom}c) are reproduced by the same set of parameters (Fig. \ref{zoom}d), using  Eq.\ \ref{S21-cav2} and Eq.\ \ref{dS21dTpha}.

\subsection{Electrodynamic simulations of cavity modes}
\begin{figure}
\begin{center}
\includegraphics[width=\linewidth]{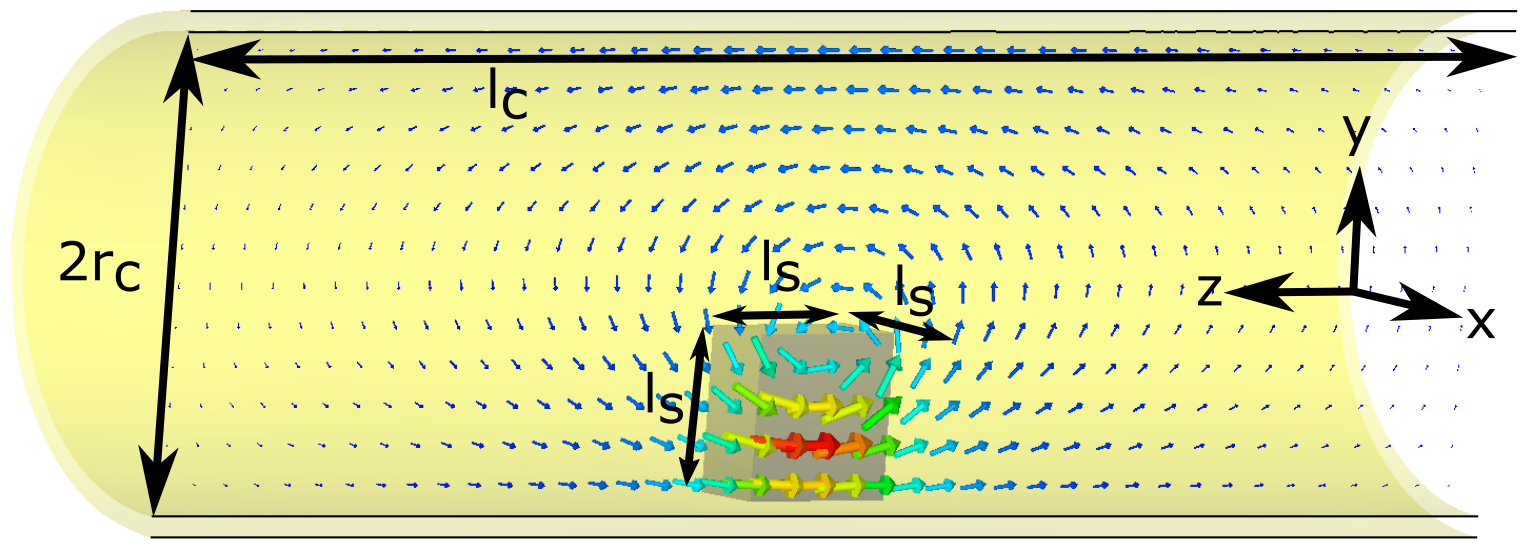}
\caption{{\label{mode-sim}
Model of the cavity drawn to-scale in cross-section. Arrows show strength of magnetic field of the 1st cavity mode on the $x$ plane.
}}
\end{center}
\end{figure}
\begin{figure}
\begin{center}
\includegraphics[width=\linewidth]{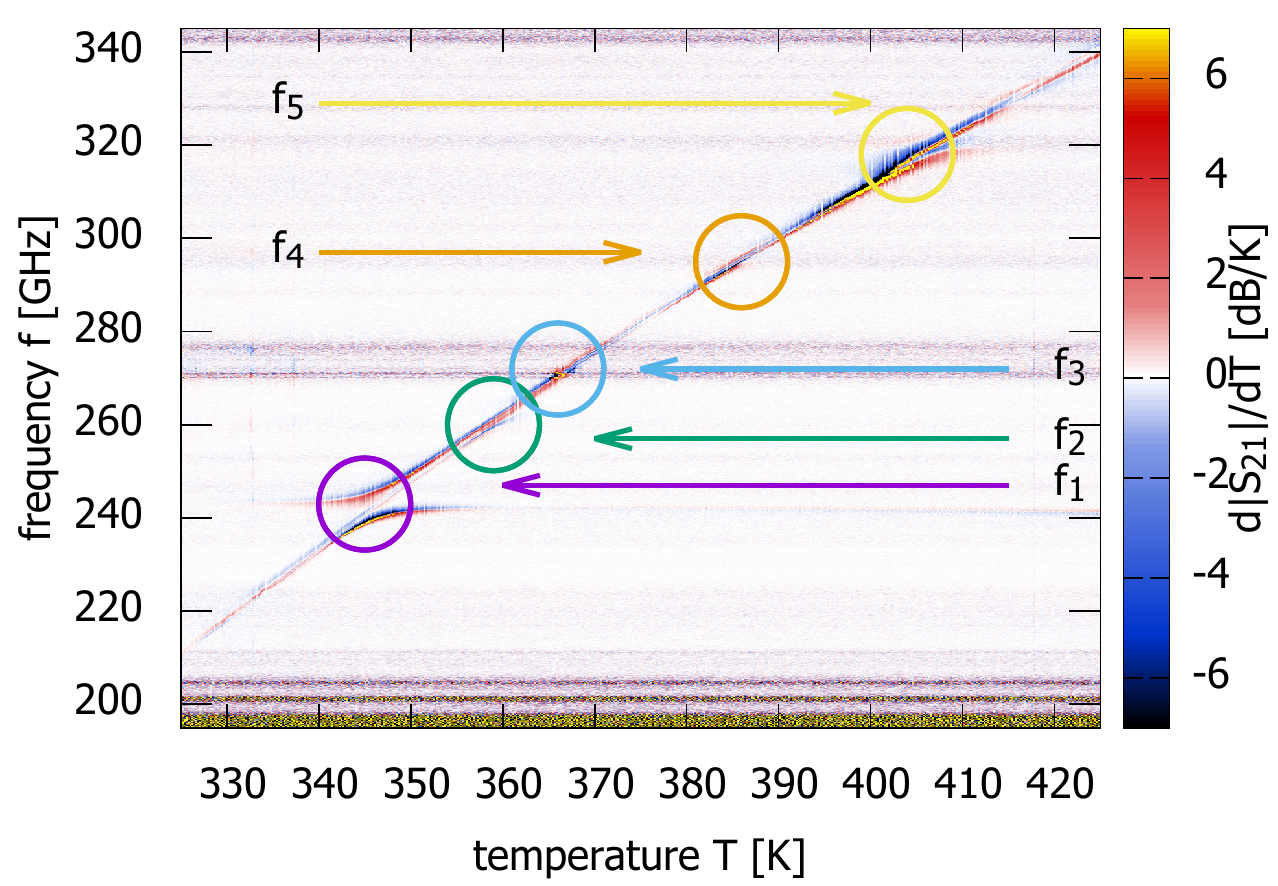}
\caption{{\label{broad}
Temperature-derivative spectra over a broad temperature range. Arrows mark expected frequencies of cavity modes and circles mark observed interactions.
}}
\end{center}
\end{figure}
We identified the cavity modes using a numerical electrodynamics field simulation software (CST Microwave Studio). We modeled the cavity as a metallic cylinder containing an isotropic dielectric cube (Fig.\ \ref{mode-sim}a). We assumed that the dielectric constant of hematite was 19.1 at 350~K, as determined from our measurements on bulk samples \cite{bulk_hematite}. We calculated the dependence of modes on the size and position of the cube inside the cylinder (see appendix).
The actual size of the cube was the smallest we could achieve by our cutting process.


We give in Tab.\ \ref{modes} the expected frequencies of the first five modes and dominant directions of the electric field in the cavity, as well as the dominant direction of the magnetic field in the cube. Only the 5th mode has a similar symmetry to that of the 1st mode, and is able to excite antiferromagnetic resonance in the sample. This suggest that the  magnetization vector  $\mathbf{m}$ in the cube was aligned along the $z$ axis of the cavity during the experiment. Under this assumption, modes 2--4 do not excite the magnetic resonance and thus, should only weakly hybridize with it. This agrees with our observation in Fig.\ \ref{broad}a, that only the 5th mode produces some strong coupling, though not as clear as the coupling to the 1st mode. This may be due to a more complex distribution of the cavity magnetic field in the cube, i.e. the ratio of $h_z/|h_z|\approx0.53$ is lower than in the case of the 1st mode (Tab.\ \ref{modes}). This means that some of the spins in the cube are excited out of phase.
\begin{table}
\centering
\begin{tabular}{ cc|c|ccc|ccc|c } 
 \hline\hline
        &             & cavity    & \multicolumn{7}{c}{mean $h$-field in the cube [arb.u.]} \\
 $j$    & $f_j$ [GHz] & $e$-field & $h_x$ & $h_y$ & $h_z$ & $|h_x|$ & $|h_y|$ & $|h_z|$ & $|\mathbf{h}|$\\
 \hline
 $1$    & $247$       & $x$ & 0.01 & -0.01 & 2.29  & 0.42 & 0.66 & \textbf{2.39} & 2.70\\
 $2$    & $258$       & $y$ & 0.12 & 0.00  & 0.00  & 0.96 & 0.09 & 1.30 & 1.76 \\
 $3$    & $272$       & $z$ & 2.51 & 0.00  & -0.01 & \textbf{2.61} & 0.75 & 0.63 & 3.05\\
 $4$    & $297$       & -   & 0.00 & 2.30  & 0.01  & 0.72 & \textbf{2.51} & 0.90 & 3.15\\
 $5$    & $329$       & $x$ & 0.00 & 0.00  & -1.00 & 0.54 & 0.43 & \textbf{1.87} & 2.19\\
 \hline\hline
\end{tabular}
\caption{\label{modes}Predicted modes frequencies, selection rules and mean magnetic field in the cube.}
\end{table}


\section{Summary}
In conclusion, we observed strong coupling between the quasi-antiferromagnetic resonance (qAFMR) of a hematite cube ($\alpha$-Fe$_2$O$_3$) located in a cavity near room temperature. 
We showed a method of achieving a strong light-matter coupling regime in antiferromagnets which is applicable to a broad range of these materials
The avoided crossing, which is characteristic of polariton dynamics, occurred near 243~GHz. The cooperativity $C$, which is a measure of the coupling strength with respect to both the magnetic resonance and the cavity line widths (Eq. \ref{coop_factor}), was estimated at 40. 

\begin{acknowledgments}
We would like to thank Claude Amendola for help in fabricating samples. 
Support by the Sino-Swiss Science and Technology Cooperation (SSSTC) grant no.\ EG-CN\_02\_032019 is gratefully acknowledged. The VNA and frequency extenders were funded by EPFL and the SNF R'Equip under Grant No.\ 206021\_144983.
\end{acknowledgments}

\appendix*
\section{A: Temperature dependence of the antiferromagnetic resonance and the 1st cavity mode}
\begin{figure}[h]
\begin{center}
\includegraphics[width=\linewidth]{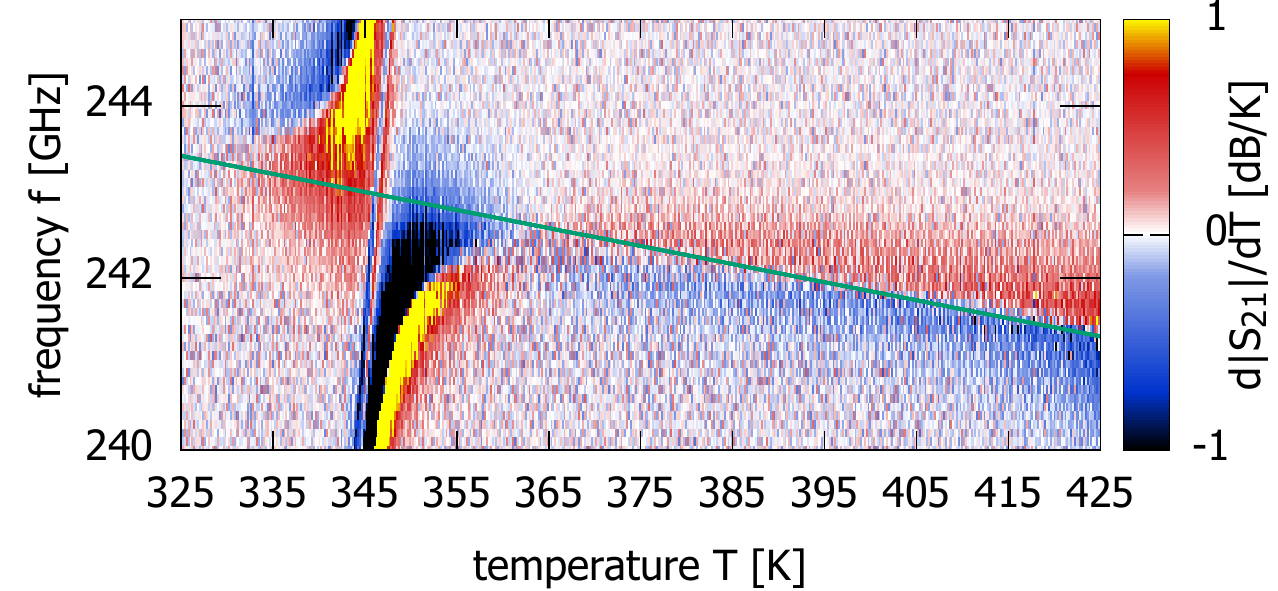}
\caption{{\label{broad-appendix}
Close-up of the first cavity mode in an over-saturated scale. The green line shows the assumed linear temperature dependence of the first mode frequency.
}}
\end{center}
\end{figure}
Fig.\ \ref{broad-appendix} shows the observed temperature dependence of the 1st mode, that we determined as
\begin{equation*}
   f_1(T)=f_1^{345} + f_1^T(T-T_1),
\end{equation*}
were $f_1^{345}=243$~GHz, $T_1=345$~K and $f_1^T=-2.1\times10^{-2}$~GHz/K. This dependence is caused by the dielectric constant of hematite increasing with rising temperature. For the cube, this dependence on the dielectric constant is predicted to be small (Fig.\ \ref{mode-sim-appendix}f).

In the configuration (a) (Fig.\ \ref{cav-res} and Fig.\ \ref{cav-delta}d) the temperature dependence of the antiferromagnetic was fitted with a linear function
\begin{equation*}
   f_r(T)=f_r^{330} + f_r^T(T-T_r),
\end{equation*}
were $f_r^{330}=230.7$~GHz, $T_r=330$~K and $f_r^T=1.176$~GHz/K. In the configuration (b) (Fig.\ \ref{cav-delta}fh), we fitted: $f_r^{330}=220.1$~GHz, and $f_r^T=1.410$~GHz/K. In Figs.\ \ref{zoom} and \ref{broad}, where we show  temperature dependence on a broader scale, it was necessarily to assume that the frequency of qAFMR is described by a parabolic function
\begin{equation*}
   f_r(T)=f_r^{330} + f_r^T(T-T_r) + f_r^{T2}(T-T_r)^2,
\end{equation*}
where $f_r^{330}=218.4$~GHz, $T_r=330$~K, $f_r^T=1.596$~GHz/K and $f_r^{T2}=-5.15\times10^{-3}$~GHz/K$^2$.
These differences of $f_r(T)$ result from different placements of the thermocouple in the holder, and thus, a different temperature calibration.

\appendix*
\section{B: Frequency dependence of modes on cavity geometry}
\begin{figure}
\begin{center}
\includegraphics[width=\linewidth]{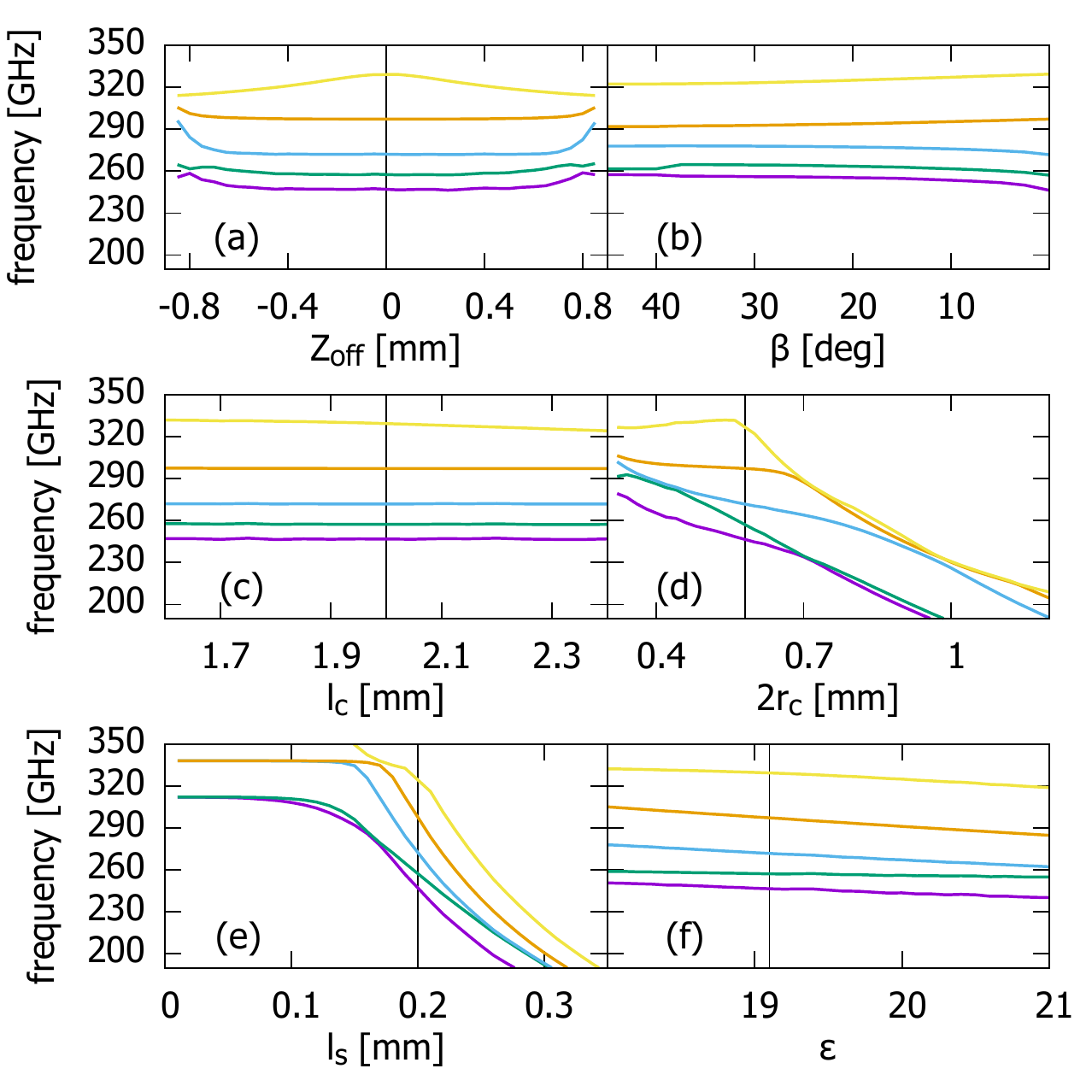}
\caption{{\label{mode-sim-appendix}
Dependence of the modes frequencies on (a) displacement of the cube from the center of the tube, (b) rotation of the cube about $y$ axis, (c) length of the tube, (d) diameter of the tube, (e) cube edge length, (f) dielectric constant of the cube. In (a)(c)(d)(e) and (f) vertical lines mark nominal values of these parameters.
}}
\end{center}
\end{figure}
The size of the cube $l_s$ was limited by our cutting process. At this smallest size, the mode frequencies are governed mostly by the size of the cube $l_s$ (Fig.\ \ref{mode-sim-appendix}e) and, to a smaller extend, by the diameter of the tube $2r_c$ (Fig.\ \ref{mode-sim-appendix}d). The length of the tube $l_c$ is irrelevant in the range of about 1.5--2.5~mm (Fig.\ \ref{mode-sim-appendix}c). For a given cube size, we chose such a diameter of the tube that the 1st and 2nd modes are well split (Fig.\ \ref{mode-sim-appendix}d). We found, that a random position of the cube in the needle spans a mode frequency range of at most 15~GHz (Fig.\ \ref{mode-sim-appendix}ab). In other calculations (Fig.\ \ref{mode-sim-appendix}c--f), the cube was assumed to be in the center of the needle ($Z_{off}=0$) and edges parallel to its axis ($\beta=0$). The calculated frequency of the 1st mode is a few GHz higher than the observed one. This might be due to inaccuracy in the cube dimensions. We observed the biggest discrepancy with the predicted and observed frequencies of the 5th mode (Fig.\ \ref{broad}). This could be related to the fact that it is the mode which shows the strongest dependence on displacement of the cube from the cavity center (Fig.\ \ref{mode-sim-appendix}a).

\begin{figure}
\begin{center}
\includegraphics[width=\linewidth]{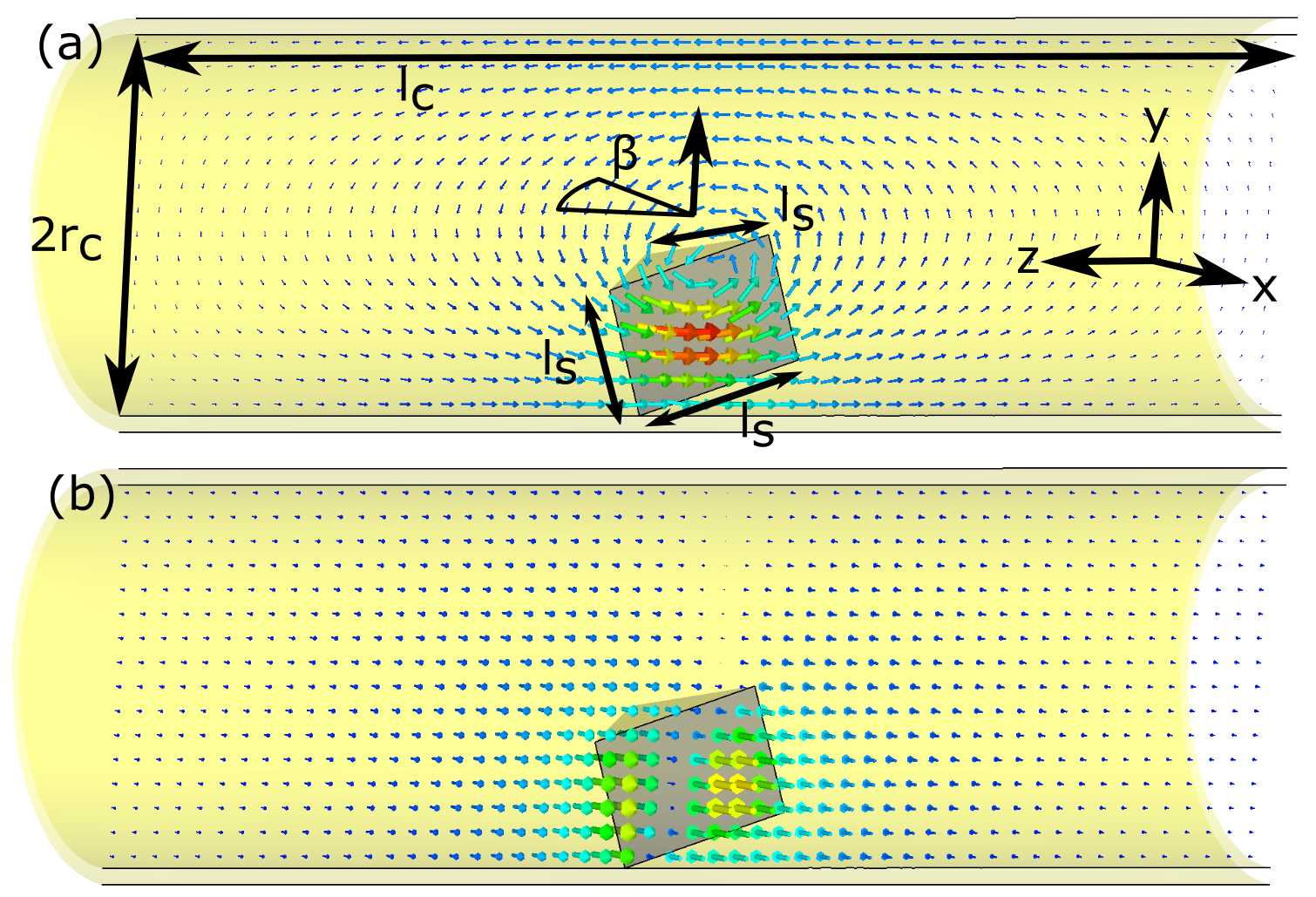}
\caption{{\label{mode-sim45}
(a) Model of the cavity drawn to-scale in cross-section. The cube is rotated about the $y$ axis by an angle $\beta=45^{\circ}$. Rotation about $x$ axis is bound by the condition of the cube touching the tube by its three corners. Arrows show direction and strength of magnetic field of the 1st cavity mode on the $x$ plane. (b) Field in the second mode the cavity on the $x$ plane.
}}
\end{center}
\end{figure}

\begin{table}
\centering
\begin{tabular}{ cc|c|ccc|ccc|c } 
 \hline\hline
        &             & cavity    & \multicolumn{7}{c}{mean $h$-field in the cube [arb.u.]} \\
 $j$    & $f_j$ [GHz] & $e$-field & $h_x$ & $h_y$ & $h_z$ & $|h_x|$ & $|h_y|$ & $|h_z|$ & $|\mathbf{h}|$\\
 \hline
 $1$    & $258$       & $x$ & 0.00 & -0.02 & 2.20  & 0.43 & 0.62 & \textbf{2.32} & 2.62\\
 $2$    & $262$       & $y$ & 0.33 & 0.00  & 0.01  & 0.93 & 0.31 & 1.24 & 1.74 \\
 $3$    & $278$       & $z$ & -2.55& 0.00  & 0.00 & \textbf{2.68} & 0.63 & 0.67 & 3.06\\
 $4$    & $292$       & -   & 0.00 & 2.45  & 0.00  & 0.68 & \textbf{2.64} & 0.77 & 3.15\\
 $5$    & $322$       & $x$ & 0.00 & 0.00  & 1.29 & 0.58 & 0.69 & \textbf{2.28} & 2.71\\
 \hline\hline
\end{tabular}
\caption{\label{modes45}Predicted modes frequencies for the cube rotated by $\beta=45^{\circ}$ (Fig.\ \ref{mode-sim45}a). Selection rules and mean magnetic field in the cube are very close to that of $\beta=0^{\circ}$ (Tab.\ 1).}
\end{table}

\appendix*
\section{C: 2-port analysis}

Let us represent a cavity as a device with 2 ports. At each port, we distinguish incoming and outgoing waves. The total voltage and current at each port are given by
\begin{equation}
V_i = V_i^+ + V_i^- \qquad I_i = I_i^+ + I_i^- \qquad (i= 1, 2)
\end{equation}
One way of characterizing the cavity is to specify the parameters $A, B, C, D$ defined by
\begin{equation}
\begin{pmatrix}
V_1\\
I_1
\end{pmatrix}
=\begin{pmatrix}
A & B\\
C & D
\end{pmatrix}
\begin{pmatrix}
V_2\\
I_2
\end{pmatrix}
\end{equation}
For a two-port device in which each port sees the common impedance $Z_3$, the $A, B, C, D$ 
parameters are given by \cite{Pozar11}
\begin{equation}
A= 1 \quad B = 0 \quad C= \frac{1}{Z_3} \quad D= 1
\end{equation}
The scattering matrix is defined by 
\begin{equation}
\begin{pmatrix}
V_1^-\\
V_2^-
\end{pmatrix}
=\begin{pmatrix}
S_{11} & S_{12}\\
S_{21} & S_{22}
\end{pmatrix}
\begin{pmatrix}
V_1^+\\
V_2^+
\end{pmatrix}
\end{equation}
In particular, the $S_{21}$ parameter is given by
\begin{equation}
S_{21}=\left . \frac{V_2^-}{V_1^+}\right\vert_{V_2^+=0}
\end{equation}
It is straightforward to calculate that,
\begin{equation}\label{S21ABCD}
S_{21}= \frac{1}{1+Z_0/(2Z_3)}
\end{equation}
where $Z_0$ is the transmission line (or wave guide) characteristic impedance at the ports. 

Let us apply this formalism to analyze a magnetic material located inside the cavity. We represent the cavity by discrete $R, L, C$ components in series. We assume that the measurement setup is such that away from a resonance of the cavity or from a resonance of the material in it, the signal propagates through the cavity without loss, i.e. $S_{21}=1$. We have,
\begin{equation}\label{Z_RLC_series}
Z_3= i \omega L + \frac{1}{i \omega C} + R
\end{equation}
We  assume that the $RLC$ circuit is designed so that the transmission is nearly one at all frequencies, i.e. $Z_3 \gg Z_0$. Then \eqref{S21ABCD} yields,
\begin{equation}\label{S21_simple}
S_{21} \approx 1- \frac{Z_0}{2Z_3}
\end{equation}

We take into account the presence of the material in the cavity by assuming that the inductance $L$ is filled entirely with a medium of an effective permittivity $\mu$ written in the form of a Lorentzian line shape, so as to account for the magnetic resonance in the material 
\begin{equation}\label{Lvalue}
L=\mu L_0 = (1+\frac{\Delta \mu \, \omega_m^2}{\omega_m^2-\omega^2-j\omega\gamma_m})L_0,
\end{equation}
where $L_0$ is an inductance of an empty cavity, $\Delta \mu$ is a unitless parameter that characterizes the coupling of the material to the cavity, $\omega_m$ is its frequency and $\gamma_m$ defines its width. Using $Z_3$ given by \eqref{Z_RLC_series} and \eqref{Lvalue}, \eqref{S21_simple} gives,
\begin{equation}
S_{21}= 1- \frac{i\omega Z_0 C/2}{1+ i\omega RC - \omega^2CL_0\left( 1+\frac{\Delta \mu \, \omega_m^2}{\omega_m^2-\omega^2-i\omega\gamma_m}\right)}
\end{equation}
Let us call $\omega_c^2= 1/(L_0C)$ in reference to the resonance of the empty cavity. We consider a situation when $\omega$, $\omega_c$ and $\omega_m$ are close, so we can approximate
\begin{equation}
\omega^2-\omega_c^2= (\omega-\omega_c)(\omega+\omega_c)\approx 2\omega_c(\omega-\omega_c)
\end{equation}
Since the resonances are sharp, $\omega$ can be replaced by $\omega_c$ when $\omega$ is not in a term that leads to a divergence (in the absence of damping). Thus, we can write as in \cite{Schuster10}
\begin{equation}
S_{21}= 1+\frac{a}{i(\omega-\omega_c)-\dfrac{\kappa}{2}+ \dfrac{G^2}{i(\omega-\omega_m)- \gamma_m/2}}
\end{equation}
where $a= \omega_c^2Z_0 C/4$, $G^2= \omega_c\omega_m \Delta\mu/4$ and $\kappa= \omega_c^2RC$. 

\bibliographystyle{unsrt}
\bibliography{refs}

\begin{thebibliography}{10}

\bibitem{Torma14}
P.~Törmä and W.~L. Barnes.
\newblock Strong coupling between surface plasmon polaritons and emitters: a
  review.
\newblock {\em Reports on Progress in Physics}, 78(1):013901, dec 2014.

\bibitem{Rempe87}
G.~Rempe, H.~Walther, and N.~Klein.
\newblock Observation of quantum collapse and revival in a one-atom maser.
\newblock {\em Phys. Rev. Lett.}, 58:353--356, Jan 1987.

\bibitem{Khitrova06}
G.~Khitrova, H.~M. Gibbs, M.~Kira, S.~W. Koch, and A.~Scherer.
\newblock Vacuum rabi splitting in semiconductors.
\newblock {\em Nature Physics}, 2(2):81--90, Feb 2006.

\bibitem{Raizen89}
M.~G. Raizen, R.~J. Thompson, R.~J. Brecha, H.~J. Kimble, and H.~J. Carmichael.
\newblock Normal-mode splitting and linewidth averaging for two-state atoms in
  an optical cavity.
\newblock {\em Phys. Rev. Lett.}, 63:240--243, Jul 1989.

\bibitem{Kasprzak06}
J.~Kasprzak, M.~Richard, S.~Kundermann, A.~Baas, P.~Jeambrun, J.~M.~J. Keeling,
  F.~M. Marchetti, M.~H. Szyma{\'{n}}ska, R.~Andr{\'e}, J.~L. Staehli,
  V.~Savona, P.~B. Littlewood, B.~Deveaud, and Le~Si Dang.
\newblock Bose--einstein condensation of exciton polaritons.
\newblock {\em Nature}, 443(7110):409--414, Sep 2006.

\bibitem{Awschalom07}
D.~D. Awschalom and M.~E. Flatt{\'e}.
\newblock Challenges for semiconductor spintronics.
\newblock {\em Nature Physics}, 3(3):153--159, Mar 2007.

\bibitem{Dovzhenko18}
D.~S. Dovzhenko, S.~V. Ryabchuk, Yu.~P. Rakovich, and I.~R. Nabiev.
\newblock Light–matter interaction in the strong coupling regime:
  configurations{,} conditions{,} and applications.
\newblock {\em Nanoscale}, 10:3589--3605, 2018.

\bibitem{Kockum19}
A.~F. Kockum, A.~Miranowicz, S.~De~Liberato, S.~Savasta, and F.~Nori.
\newblock Ultrastrong coupling between light and matter.
\newblock {\em Nature Reviews Physics}, 1(1):19--40, Jan 2019.

\bibitem{Roux20}
K.~Roux, H.~Konishi, V.~Helson, and J.-Ph. Brantut.
\newblock Strongly correlated fermions strongly coupled to light.
\newblock {\em Nature Communications}, 11(1):2974, Jun 2020.

\bibitem{Schuster10}
D.~I. Schuster, A.~P. Sears, E.~Ginossar, L.~DiCarlo, L.~Frunzio, J.~J.~L.
  Morton, H.~Wu, G.~A.~D. Briggs, B.~B. Buckley, D.~D. Awschalom, and R.~J.
  Schoelkopf.
\newblock High-cooperativity coupling of electron-spin ensembles to
  superconducting cavities.
\newblock {\em Phys. Rev. Lett.}, 105:140501, Sep 2010.

\bibitem{Abe11}
E.~Abe, H~Wu, A.~Ardavan, and J.~J.~L. Morton.
\newblock Electron spin ensemble strongly coupled to a three-dimensional
  microwave cavity.
\newblock {\em Applied Physics Letters}, 98(25):251108, 2011.

\bibitem{Huebl13}
H.~Huebl, Ch.~W. Zollitsch, J.~Lotze, F.~Hocke, M.~Greifenstein, Ac. Marx,
  R.~Gross, and S.~T.~B. Goennenwein.
\newblock High cooperativity in coupled microwave resonator ferrimagnetic
  insulator hybrids.
\newblock {\em Phys. Rev. Lett.}, 111:127003, Sep 2013.

\bibitem{Zhang14}
X.~Zhang, Ch.-L. Zou, L.~Jiang, and H.~X. Tang.
\newblock Strongly coupled magnons and cavity microwave photons.
\newblock {\em Phys. Rev. Lett.}, 113:156401, Oct 2014.

\bibitem{Tabuchi14}
Y.~Tabuchi, S.~Ishino, T.~Ishikawa, R.~Yamazaki, K.~Usami, and Y.~Nakamura.
\newblock Hybridizing ferromagnetic magnons and microwave photons in the
  quantum limit.
\newblock {\em Phys. Rev. Lett.}, 113:083603, Aug 2014.

\bibitem{Tabuchi15}
Y.~Tabuchi, S.~Ishino, A.~Noguchi, T.~Ishikawa, R.~Yamazaki, K.~Usami, and
  Y.~Nakamura.
\newblock Coherent coupling between a ferromagnetic magnon and a
  superconducting qubit.
\newblock {\em Science}, 349(6246):405--408, 2015.

\bibitem{Zhang15}
X.~Zhang, Ch.-L. Zou, Na~Zhu, F.~Marquardt, L.~Jiang, and H.~X. Tang.
\newblock Magnon dark modes and gradient memory.
\newblock {\em Nature Communications}, 6(1):8914, Nov 2015.

\bibitem{Zhang16}
X.~Zhang, Na~Zhu, Ch.-L. Zou, and H.~X. Tang.
\newblock Optomagnonic whispering gallery microresonators.
\newblock {\em Phys. Rev. Lett.}, 117:123605, Sep 2016.

\bibitem{Li19}
Y.~Li, T.~Polakovic, Y.-L. Wang, J.~Xu, S.~Lendinez, Zh. Zhang, J.~Ding,
  T.~Khaire, H.~Saglam, R.~Divan, J.~Pearson, W.-K. Kwok, Zh. Xiao, V.~Novosad,
  A.~Hoffmann, and W.~Zhang.
\newblock Strong coupling between magnons and microwave photons in on-chip
  ferromagnet-superconductor thin-film devices.
\newblock {\em Phys. Rev. Lett.}, 123:107701, Sep 2019.

\bibitem{Potts20}
C.~A. Potts and J.~P. Davis.
\newblock Strong magnon–photon coupling within a tunable cryogenic microwave
  cavity.
\newblock {\em Applied Physics Letters}, 116(26):263503, 2020.

\bibitem{Lachance-Quirion20}
D.~Lachance-Quirion, S.~P. Wolski, Y.~Tabuchi, Sh. Kono, K.~Usami, and
  Y.~Nakamura.
\newblock Entanglement-based single-shot detection of a single magnon with a
  superconducting qubit.
\newblock {\em Science}, 367(6476):425--428, 2020.

\bibitem{Li20JAP}
Y.~Li, W.~Zhang, V.~Tyberkevych, W.-K. Kwok, A.~Hoffmann, and V.~Novosad.
\newblock Hybrid magnonics: Physics, circuits, and applications for coherent
  information processing.
\newblock {\em Journal of Applied Physics}, 128(13):130902, 2020.

\bibitem{Bhoi21}
B.~Bhoi, S.-H. Jang, B.~Kim, and S.-K. Kim.
\newblock Broadband photon–magnon coupling using arrays of photon resonators.
\newblock {\em Journal of Applied Physics}, 129(8):083904, 2021.

\bibitem{Niemczyk10}
T.~Niemczyk, F.~Deppe, H.~Huebl, E.~P. Menzel, F.~Hocke, M.~J. Schwarz, J.~J.
  Garcia-Ripoll, D.~Zueco, T.~H{\"u}mmer, E.~Solano, A.~Marx, and R.~Gross.
\newblock Circuit quantum electrodynamics in the ultrastrong-coupling regime.
\newblock {\em Nature Physics}, 6(10):772--776, Oct 2010.

\bibitem{Jungwirth18}
T.~Jungwirth, J.~Sinova, A.~Manchon, X.~Marti, J.~Wunderlich, and C.~Felser.
\newblock The multiple directions of antiferromagnetic spintronics.
\newblock {\em Nature Physics}, 14(3):200--203, Mar 2018.

\bibitem{Li20Nature}
J.~Li, C.~B. Wilson, R.~Cheng, M.~Lohmann, M.~Kavand, W.~Yuan, M.~Aldosary,
  N.~Agladze, P.~Wei, M.~S. Sherwin, and J.~Shi.
\newblock Spin current from sub-terahertz-generated antiferromagnetic magnons.
\newblock {\em Nature}, 578(7793):70--74, Feb 2020.

\bibitem{Li20PRL}
J.~Li, H.~T. Simensen, D.~Reitz, Q.~Sun, W.~Yuan, Ch. Li, Y.~Tserkovnyak,
  A.~Brataas, and J.~Shi.
\newblock Observation of magnon polarons in a uniaxial antiferromagnetic
  insulator.
\newblock {\em Phys. Rev. Lett.}, 125:217201, Nov 2020.

\bibitem{Reitz20}
D.~Reitz, J.~Li, W.~Yuan, J.~Shi, and Y.~Tserkovnyak.
\newblock Spin seebeck effect near the antiferromagnetic spin-flop transition.
\newblock {\em Phys. Rev. B}, 102:020408, Jul 2020.

\bibitem{Ghosh21}
A.~Ghosh, M.~Palit, S.~Maity, V.~Dwij, S.~Rana, and S.~Datta.
\newblock Spin-phonon coupling and magnon scattering in few-layer
  antiferromagnetic ${\mathrm{feps}}_{3}$.
\newblock {\em Phys. Rev. B}, 103:064431, Feb 2021.

\bibitem{Hoffman15}
A.~Hoffmann and S.~D. Bader.
\newblock Opportunities at the frontiers of spintronics.
\newblock {\em Phys. Rev. Applied}, 4:047001, Oct 2015.

\bibitem{Li11}
J.~Li, T.~Higuchi, N.~Kanda, K.~Konishi, S.~G. Tikhodeev, and
  M.~Kuwata-Gonokami.
\newblock Control of magnetic dipole terahertz radiation by cavity-based phase
  modulation.
\newblock {\em Opt. Express}, 19(23):22550--22556, Nov 2011.

\bibitem{Bialek20}
M.~Bia\l{}ek, A.~Magrez, and J.-Ph. Ansermet.
\newblock {Spin-wave coupling to electromagnetic cavity fields in dysposium
  ferrite}.
\newblock {\em Phys. Rev. B}, 101:024405, Jan 2020.

\bibitem{Everts20}
J.~R. Everts, G.~G.~G. King, N.~J. Lambert, S.~Kocsis, S.~Rogge, and J.~J.
  Longdell.
\newblock Ultrastrong coupling between a microwave resonator and
  antiferromagnetic resonances of rare-earth ion spins.
\newblock {\em Phys. Rev. B}, 101:214414, Jun 2020.

\bibitem{Grishunin18}
K.~Grishunin, T.~Huisman, G.~Q Li, E.~Mishina, T.~Rasing, A.~V. Kimel, K.L
  Zhang, Z.~M Jin, S.~X Cao, W~Ren, G.~H Ma, and Rostislav~V. Mikhaylovskiy.
\newblock {Terahertz Magnon-Polaritons in TmFeO$_{3}$}.
\newblock {\em ACS Photonics}, 5(4):1375--1380, 2018.

\bibitem{Shi20}
L.~Y. Shi, D.~Wu, Z.~X. Wang, T.~Lin, C.~M. Hu, and N.~L. Wang.
\newblock {Revealing ultra-strong magnon-photon coupling in a polar
  antiferromagnet Fe$_2$Mo$_3$O$_8$ by time domain terahertz spectroscopy},
  2020.

\bibitem{Li18}
X.~Li, M.~Bamba, N.~Yuan, Qi~Zhang, Y.~Zhao, M.~Xiang, K.~Xu, Z.~Jin, W.~Ren,
  G.~Ma, Sh. Cao, D.~Turchinovich, and J.~Kono.
\newblock Observation of dicke cooperativity in magnetic interactions.
\newblock {\em Science}, 361(6404):794--797, 2018.

\bibitem{Sivarajah19}
P.~Sivarajah, A.~Steinbacher, B.~Dastrup, J.~Lu, M.~Xiang, W.~Ren, S.~Kamba,
  Sh. Cao, and K.~A. Nelson.
\newblock {THz-frequency magnon-phonon-polaritons in the collective
  strong-coupling regime}.
\newblock {\em Journal of Applied Physics}, 125(21):213103, 2019.

\bibitem{Hu20}
C.-M. Hu.
\newblock Chapter four - the 2020 roadmap for spin cavitronics.
\newblock volume~71 of {\em Solid State Physics}, pages 117 -- 121. Academic
  Press, 2020.

\bibitem{Przenioslo14}
R.~Przeniosło, I.~Sosnowska, M.~Stekiel, D.~Wardecki, A.~Fitch, and J.~B.
  Jasiński.
\newblock {Monoclinic deformation of the crystal lattice of hematite
  $\alpha$-Fe$_2$O$_3$}.
\newblock {\em Physica B: Condensed Matter}, 449:72 -- 76, 2014.

\bibitem{Morin50}
F.~J. Morin.
\newblock {Magnetic Susceptibility of
  $\ensuremath{\alpha}{\mathrm{Fe}}_{2}{\mathrm{O}}_{3}$ and
  $\ensuremath{\alpha}{\mathrm{Fe}}_{2}{\mathrm{O}}_{3}$ with Added Titanium}.
\newblock {\em Phys. Rev.}, 78:819--820, Jun 1950.

\bibitem{Aleksandrov85}
K.~S. Aleksandrov, L.~N. Bezmaternykh, G.~V. Kozlov, S.~P. Lebedev, A.~A.
  Mukhin, and A.~S. Prokhorov.
\newblock {Anomalies of high-frequency magnetic permeability of hematite at the
  Morin phase transition}.
\newblock {\em Journal of Experimental and Theoretical Physics}, 65(3):591,
  1986.

\bibitem{DZYALOSHINSKY58}
I.~Dzyaloshinsky.
\newblock A thermodynamic theory of “weak” ferromagnetism of
  antiferromagnetics.
\newblock {\em Journal of Physics and Chemistry of Solids}, 4(4):241 -- 255,
  1958.

\bibitem{Caspers16}
C.~Caspers, V.~P. Gandhi, A.~Magrez, E.~de~Rijk, and J.-P. Ansermet.
\newblock {Sub-terahertz spectroscopy of magnetic resonance in BiFeO$_3$ using
  a vector network analyzer}.
\newblock {\em Applied Physics Letters}, 108(24):241109, 2016.

\bibitem{Bialek18}
M.~Bia\l{}ek, A.~Magrez, A.~Murk, and J.-Ph. Ansermet.
\newblock Spin-wave resonances in bismuth orthoferrite at high temperatures.
\newblock {\em Phys. Rev. B}, 97:054410, Feb 2018.

\bibitem{Bialek19}
M.~Bia\l{}ek, T.~Ito, H.~R\o{}nnow, and J.-Ph. Ansermet.
\newblock Terahertz-optical properties of a bismuth ferrite single crystal.
\newblock {\em Phys. Rev. B}, 99:064429, Feb 2019.

\bibitem{Zhang20}
J.~Zhang, M.~Białek, A.~Magrez, H.~Yu, and J.-Ph. Ansermet.
\newblock {Antiferromagnetic resonance in TmFeO$_3$ at high temperatures}.
\newblock {\em Journal of Magnetism and Magnetic Materials}, 523:167562, 2021.

\bibitem{Harder16}
M.~Harder, L.-H. Bai, Ch. Match, J.~Sirker, and C.-M. Hu.
\newblock Study of the cavity-magnon-polariton transmission line shape.
\newblock {\em Science China Physics, Mechanics {\&} Astronomy}, 59(11):117511,
  Sep 2016.

\bibitem{Mills74}
D~L Mills and E~Burstein.
\newblock Polaritons: the electromagnetic modes of media.
\newblock {\em Reports on Progress in Physics}, 37(7):817--926, jul 1974.

\bibitem{Soykal10}
\"O.~O. Soykal and M.~E. Flatt\'e.
\newblock Strong field interactions between a nanomagnet and a photonic cavity.
\newblock {\em Phys. Rev. Lett.}, 104:077202, Feb 2010.

\bibitem{Niemczyk09}
T~Niemczyk, F~Deppe, M~Mariantoni, E~P Menzel, E~Hoffmann, G~Wild,
  L~Eggenstein, A~Marx, and R~Gross.
\newblock Fabrication technology of and symmetry breaking in superconducting
  quantum circuits.
\newblock {\em Superconductor Science and Technology}, 22(3):034009, jan 2009.

\bibitem{Flower19}
G.\ Flower, M.\ Goryachev, J.\ Bourhill, and M.~E. Tobar.
\newblock Experimental implementations of cavity-magnon systems: from ultra
  strong coupling to applications in precision measurement.
\newblock {\em New Journal of Physics}, 21(9):095004, sep 2019.

\bibitem{Bourhill20}
J.~Bourhill, V.~Castel, A.~Manchec, and G.~Cochet.
\newblock Universal characterization of cavity–magnon polariton coupling
  strength verified in modifiable microwave cavity.
\newblock {\em Journal of Applied Physics}, 128(7):073904, 2020.

\bibitem{Pailhe08}
N.~Pailh\'e, J.~Majimel, S.~Pechev, P.~Gravereau, M.~Gaudon, and A.~Demourgues.
\newblock {Investigation of Nanocrystallized $\alpha$-Fe$_2$O$_3$ Prepared by a
  Precipitation Process}.
\newblock {\em The Journal of Physical Chemistry C}, 112(49):19217--19223,
  2008.

\bibitem{Shull51}
C.~G. Shull, W.~A. Strauser, and E.~O. Wollan.
\newblock Neutron diffraction by paramagnetic and antiferromagnetic substances.
\newblock {\em Phys. Rev.}, 83:333--345, Jul 1951.

\bibitem{bulk_hematite}
M.~Bia\l{}ek, J.~Zhang, H.~Yu, and J.-Ph. Ansermet.
\newblock {in preparation}.

\bibitem{Pozar11}
D.~M. Pozar.
\newblock {\em Microwave engineering}.
\newblock John Wiley \& sons, 2011.

\end{thebibliography}

\end{document}